\documentclass[aps,prd,twocolumn,preprintnumbers,superscriptaddress]{revtex4-1}
\usepackage{graphicx}
\usepackage{fp}
\usepackage{epstopdf}
\usepackage{amsmath}
\usepackage{amsfonts}
\usepackage{amssymb}
\usepackage{appendix}
\usepackage{enumerate}
\usepackage{natbib}
\usepackage{comment}
\usepackage{bbold}
\usepackage[shortlabels]{enumitem}
\usepackage{color}
\usepackage{slashed}
\usepackage{subfigure}
\usepackage{setspace}
\usepackage{footnote}
\usepackage{lipsum}
\usepackage{multirow}
\usepackage[colorlinks = true,
            linkcolor = blue,
            urlcolor  = blue,
            citecolor = blue,
            anchorcolor = blue]{hyperref}
\usepackage[capitalize]{cleveref}
\usepackage{braket}
\usepackage[compat=1.1.0]{tikz-feynman}
\usepackage{multirow}
\usepackage{physics}
\usepackage{feynmp-auto}
\usepackage[normalem]{ulem}
\usepackage{url}
\usepackage{units}
\usepackage[normalem]{ulem}

\newcommand{\be}{\begin{eqnarray}}
\newcommand{\ee}{\end{eqnarray}}
\newcommand{\ba} {\begin{equation}\begin{aligned}}
\newcommand{\ea} {\end{aligned}\end{equation}}
\newcommand{\bg} {\begin{equation}\begin{gathered}}
\newcommand{\eg} {\end{gathered}\end{equation}}

\newcommand{\beq}{\begin{equation}}
\newcommand{\eeq}{\end{equation}}

\newcommand{\MeV}{\text{MeV}}

\newcommand{\seg}{\ \text{s}}

\newcommand{\sL}{\mathcal{L}}

\newcount\boxheight
\newcount\boxwidth
\newcommand\testaspect[1]{%
  \setbox0=\hbox{#1}%
  \boxheight=\ht0\relax%
  \boxwidth=\wd0\relax%
  \FPdiv\theaspect{\the\boxheight}{\the\boxwidth}%
  \copy0%
}

\begin{document}

\bigskip\
\preprint{CERN-TH-2023-178}

\title{New Directions for ALP Searches Combining Nuclear Reactors and Haloscopes}
\author{Fernando~Arias-Arag\'on}
\affiliation{Istituto Nazionale di Fisica Nucleare, Laboratori Nazionali di Frascati, C.P. 13, 00044 Frascati, Italy}
\affiliation{Laboratoire de Physique Subatomique et de Cosmologie, Universit\'e Grenoble-Alpes, CNRS/IN2P3, Grenoble INP, 38000 Grenoble, France}
\author{Vedran~Brdar}
\affiliation{CERN, Theoretical Physics Department, 1211 Geneva 23, Switzerland}
\affiliation{Department of Physics, Oklahoma State University, Stillwater, OK, 74078, USA}
\author{J\'er\'emie~Quevillon}
\affiliation{Laboratoire de Physique Subatomique et de Cosmologie, Universit\'e Grenoble-Alpes, CNRS/IN2P3, Grenoble INP, 38000 Grenoble, France}
\affiliation{CERN, Theoretical Physics Department, 1211 Geneva 23, Switzerland}

\begin{abstract}
In this work we propose reactoscope, a novel experimental setup for axion-like particle (ALP) searches. Nuclear reactors produce a copious number of photons, a fraction of which could convert into ALPs via Primakoff process in the reactor core. The generated flux of ALPs leaves the nuclear power plant and its passage through a region with a strong magnetic field results in the efficient conversion to photons which can be detected. Such magnetic field is the key component of axion haloscope experiments. Adjacent nuclear reactor and axion haloscope experiment exist in Grenoble, France. There, the Institut Laue–Langevin (ILL) research reactor is situated only $\sim 700$ m from GrAHal, the axion haloscope platform designed to offer several volume and magnetic field (up to 43 T) configurations. We derive sensitivity projections for photophilic ALP searches with ILL and GrAHal, and also scrutinize analogous realizations, such as the one comprising of CAST experiment at CERN and Bugey nuclear power plant. The results that we obtain complement and extend the reach of existing laboratory experiments, e.g. light-shining-through-walls. While the derived sensitivities are not competitive when compared to the astrophysical limits, our analysis is free from the assumptions associated to those.  
\end{abstract}

\maketitle

\textbf{Introduction.}
The Standard Model (SM) of particle physics is an extremely successful theory that has described with astonishing accuracy most observed phenomena in high energy physics. Despite its predictive power, the model presents some issues, both at the theoretical level in the form of fine-tuned parameters and at the experimental level, with measurements directly incompatible with SM predictions. These open problems allow us to investigate new physics realizations capable of addressing the aforementioned shortcomings.

In this work we will focus on one particular beyond the Standard Model realization -- ALPs (denoted as $a$) which have properties similar to those of QCD axions with the notable difference in the fact that the particle's mass and its decay constant are treated as independent parameters. Regarding experimental efforts, see \emph{e.g.} \cite{Redondo:2010dp,Adams:2022pbo,Bauer:2017ris} for a broad summary of ALP constraints from terrestrial experiments as well as cosmological and astrophysical probes.

We will focus on the following interaction of ALPs with the SM
\begin{align}
\sL_{a\gamma\gamma}= - \frac{1}{4}\,g_{a\gamma\gamma}\,a\,F^{\mu\nu}\,\tilde{F}_{\mu\nu}\,,
\label{eq:lag}
\end{align}
where $g_{a\gamma\gamma}$ is the interaction strength (in units of $\text{GeV}^{-1}$) 
and $F^{\mu\nu}$ is the field strength tensor of the electromagnetic field, where $\tilde{F}_{\mu\nu}=\frac{1}{2}F^{\alpha\beta}\varepsilon_{\alpha\beta\mu\nu}$ 
is its dual, with $\varepsilon_{1230}=1$.

We propose a new realization, dubbed reactoscope, for testing ALPs that interact via \cref{eq:lag}. Namely, we put forward a possibility of combining two different experimental facilities: ALPs are produced in nuclear reactors via scattering process that occurs through the interaction in \cref{eq:lag} and, due to the same $\gamma$ -- $a$ coupling, ALPs can convert back to photons in a magnetic field. An experimental realization involving the strong magnetic field is provided in the haloscope experiments which utilize resonant cavities \cite{Sikivie:1983ip}. Such cavities are used for resonant conversion of light axion dark matter particles; the conversion is maximized when axion mass matches the resonant frequency \cite{Alesini:2023qed}. In contrast, the cavities are not useful within our setup since, for reactor ALPs ($\mathcal{O}(\text{MeV})$ energy), the resonant condition cannot be met. However, the strong magnetic field from such experiments is crucial for the method proposed in this work. 

The realization with the nuclear reactor and the resonant cavity experiment in the vicinity to each other exists in Grenoble, France where ILL research reactor is placed only around $700$ meters from GrAHal, the axion haloscope platform. GrAHal will be able to run detectors of different sizes and designs, with an axion and ALP mass sensitivity in the range of $1.25$ to $125\ \mu \rm{eV}$~\cite{Grenet:2021vbb,9714155,2022MS&E.1240a2122P}. One of its main components is the Grenoble hybrid magnet. This magnet can generate a magnetic field up to 43 T within a 34 mm diameter aperture. Another configuration will allow for the production of 17.5 T in a 375 mm diameter aperture and 9.5 T in an 812 mm diameter bore. Additionally, several superconducting coils capable of generating magnetic fields up to 20 T within a 50 mm diameter will allow for the simultaneous operation of multiple haloscopes.
We are proposing an experiment which should not interfere with scheduled program of neither ILL nor GrAHal. Additionally, such a realization would by no means require a large financial commitment given that almost all the required components are already operating. Additionally, we should also stress that ILL is equipped with several strong magnets that can also be utilized for axion searches; such magnets are presently used in different branches of physics, primarily involving neutrons~\cite{ILLHori,ILLVert}. A similar realization to ILL-GrAHal that will also be considered in this work is the combination of the Bugey nuclear power plant (near the city of Lyon, in France) and CAST helioscope \cite{CAST:2004gzq,CAST:2017uph} based at CERN.

Regarding the production of ALPs in nuclear reactors, let us bring up Refs.~\cite{Dent:2019ueq,AristizabalSierra:2020rom} where the authors studied detection of ALPs via scattering or decay in neutrino experiments (for other ALP searches see e.g.~\cite{Brdar:2020dpr,Kelly_2021,Kling:2022ehv,Brdar:2022vum}). In this work we are instead focused on conversion in the magnetic field and we identify regions in the parameter space that exceed sensitivity projections derived in \cite{Dent:2019ueq,AristizabalSierra:2020rom}. Furthermore, let us also point out Ref.~\cite{Bonivento:2019sri} where the authors considered ALP production in accelerator experiments and a subsequent detection through the conversion in a magnetic field, where the latter can be achieved for instance by reusing \emph{e.g.} CAST magnets. While we consider such a proposal appealing, we stress that the advantage of the setup proposed in this work is that the components for both ALP production and detection (adjacent nuclear reactor and magnetic field) are already present and operating at the aforementioned sites. 

In the remainder of the paper we focus on computing sensitivity projections for the proposed experimental setup.\\\\
\textbf{ALP production at reactors.}
Nuclear reactors are powerful source of neutrons, neutrinos and photons and it is the latter that are relevant for the production of ALPs, given \cref{eq:lag}. The photon fluxes in the reactor core vary across different reactor facilities as they depend on the fuel and the configuration of the core, which we treat as monolithic uranium. In the center of our interest is the ILL reactor for which, to the best of our knowledge, the detailed photon flux estimation is not publicly available. Hence, as a proxy, we will use the analysis for the FRJ-1 research reactor \cite{reactor} which as well used $^{235}$U for the nuclear fission. Such a strategy was also employed recently in the literature in the context of both dark photon \cite{Park:2017prx,Ge:2017mcq} and ALP \cite{AristizabalSierra:2020rom} production.    

We parametrize the photon flux, arising both from fission itself and subsequent beta decays, in the core as \cite{AristizabalSierra:2020rom,reactor}
\begin{align}
\frac{d\Phi_\gamma}{dE_\gamma}=\frac{0.58\times 10^{21}}{\MeV\, \seg} \left(\frac{P}{\text{GW}}\right)e^{-1.1\frac{E_\gamma}{\MeV}}\,,
\label{eq:gammaflux}
\end{align}
where $P$ is the thermal power of the reactor and $E_\gamma$ is the photon energy.
ILL reactor has a thermal power of $58$ MW, while the commercial power plant Bugey has a thermal power of roughly $3.6$ GW.
We note that \cref{eq:gammaflux} has uncertainties. While it is beyond the scope of this work to perform a detailed simulation for the particular reactor, we have compared this parametrization with the simulation results from Ref. \cite{Dent:2019ueq}. We found that the two integrated photon fluxes differ roughly by a factor of 2.  Given the 
$g_{a\gamma\gamma}^4$ dependence of the event rate (to be shown later), the factor of 2 uncertainty in the flux would lead only to a 20\% ($2^{0.25}\simeq 1.2$) correction in the sensitivity projections for $g_{a\gamma\gamma}$. 

Photons in the reactor core can interact with the nuclear fuel, namely $^{235}$U. In most cases, this interaction would lead to absorption or scattering of a photon; however, once in a while photon can produce an ALP via Primakoff scattering, i.e. $N+\gamma\rightarrow N+a$, where N is an atomic nucleus. 

\begin{figure}[t]
    \centering
    \includegraphics[height=160 pt,trim=1.5 1.7 1.5 1, clip]{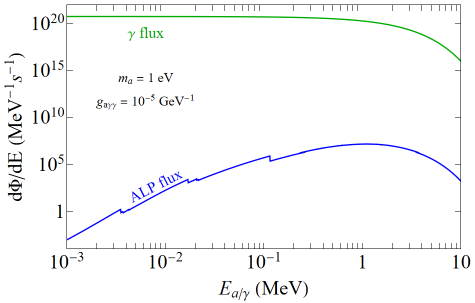}
    \caption{Comparison of the $\gamma$ flux in \cref{eq:gammaflux} and 
    ALP flux in \cref{eq:flux}. Reactor thermal power is taken as $1$ GW.}
    \label{fig:flux}
\end{figure}

The differential cross section for this process reads~\cite{Aloni:2019ruo}
\bg
\frac{d\sigma_{\text{Prim}}}{dt} = 2\,\alpha\, Z^2 \, |F(t)|^2 \,g_{a\gamma\gamma}^2\,M_N^4\  \\ \times \frac{m_a^2\, t\, (M_N^2+s)-m_a^4M_N^2-t\left(\left(M_N^2-s\right)^2+st\right)}{t^2\left(M_N^2-s\right)^2\left(t-4M_N^2\right)^2}\,,
\label{eq:cross}
\eg
where $t$ and $s$ are the Mandelstam variables, $Z$, $M_N$ and $m_a$ are the atomic number, mass of the $^{235}$U, and ALP mass, respectively. Further, $\alpha$ is the fine structure constant and $F^2(t)$ is the nuclear form factor; see \emph{e.g.} \cite{Abe:2020mcs,Tsai:1973py} for the definitions. In passing, let us also stress that Primakoff process can occur via scattering on electrons as well ($\gamma \, e^- \to a \, e^-$); however, scattering on $^{235}$U strongly dominates because of the $Z^2$ enhancement in the cross section. Notice that nuclear reactors emit photons with energies of at most $\mathcal{O}(10)$ MeV, whereas the mass of $^{235}$U is $M_N\sim 200$ GeV. This implies that the energy of the photon that scatters on nucleus and the energy of ALP that is produced in such an interaction approximately coincide; we therefore take $E_\gamma=E_a$ in order to simplify the calculations. Then, the ALP flux at the production site reads 
\begin{align}
\frac{d\Phi_a^0}{dE_a} = \frac{\sigma_{\text{Prim}}(E_a)}{\sigma_\text{tot}(E_a)}\frac{d\Phi_\gamma(E_a)}{dE_a}\,,
\label{eq:flux}
\end{align}
where $\sigma_\text{tot}$ is the total cross section for the scattering of a photon with $^{235}$U, including SM and ALP contributions. We assume that the photon is extinguished in each interaction, which implies an underestimation in the photon (and consequently ALP) flux at low  energies. This means that our calculations are conservative. Similar as in \cite{AristizabalSierra:2020rom}, we extracted $\sigma_\text{tot}$  from \cite{148746}.
In \cref{fig:flux}, we make a comparison between the photon and ALP fluxes at the production site. The latter is suppressed due to small $\sigma_{\text{Prim}}/\sigma_\text{tot}$.

The produced ALPs will travel a certain distance, and we need to take into account both their survival probability and the spatial dilution of the flux. 
Regarding the former, given the interaction in \cref{eq:lag}, ALPs can decay to two photons and we are interested in ALPs that do not decay before reaching distance $D$ from the reactor core. The decay rate reads
\begin{align}
\Gamma(a\rightarrow \gamma\gamma)=\frac{g_{a\gamma\gamma}^2}{64\pi}m_a^3\,.
\end{align}
The lifetime in the laboratory frame is hence $\tau=(E_a/m_a)\,\sqrt{1-m_a^2/E_a^2}\, \Gamma^{-1}(a\rightarrow \gamma\gamma)$ and the survival probability reads $P_{\text{surv}}=\text{Exp}[-D/(c\tau)]$.

\begin{figure}[t]
    \centering\vspace{2.2mm}
    \includegraphics[height=160 pt,trim=1 0 1 2, clip]{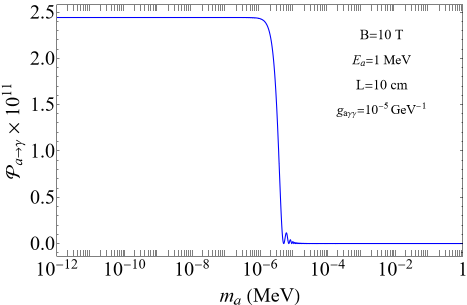}
    \caption{$\mathcal{P}_{a\rightarrow\gamma}$ as a function of $m_a$.}
    \label{fig:conv}
\end{figure}

Regarding the latter, for nuclear reactors we assume an isotropic production as a conservative estimate~\footnote{We thank Rafik Ballou for his comments about the geometry of the flux.}. Given the above, we are now able to express ALP flux at the distance $D$ from the core
\begin{align}
\frac{d\Phi_a}{dE_a} = \frac{\mathcal{P}_{\text{surv}}}{4\pi D^2}\frac{\sigma_{\text{Prim}}(E_a)}{\sigma_\text{tot}(E_a)}\frac{d\Phi_\gamma}{dE_a}\,.
\label{eq:flux2}
\end{align}\\\\
\textbf{ALP detection.}
Let us now turn our attention towards the detection. Here, we envision $a\to \gamma$ conversion in the magnetic field, in the similar spirit as it is done at helioscopes like CAST \cite{CAST:2004gzq,CAST:2017uph}. The conversion probability of an ALP, that travels across the distance $L$ immersed in a magnetic field $B$, into a photon reads \cite{Raffelt:1987im,DeAngelis:2011id,Dev:2021ofc}
\begin{align}
\mathcal{P}_{a\rightarrow\gamma}=\left(\frac{g_{a\gamma\gamma} B}{q}\right)^2 \sin^2\left(\frac{qL}{2}\right),
\label{eq:conv}
\end{align}
with $q=\sqrt{\left(m_a^2/(2 E_a)\right)^2+\left(g_{a\gamma\gamma}B\right)^2}$. For 
small $m_a$ and $g_{a\gamma\gamma}$, which will appear to represent the parameter space of our interest, it can be shown that $\mathcal{P}_{a\rightarrow\gamma}\approx (g_{a\gamma\gamma} B L/2)^2$; namely, the dependence on $q$ drops due to $\lim_{x\to 0}\, (\sin^2 x/x^2) =1$. $\mathcal{P}_{a\rightarrow\gamma}$ is shown in \cref{fig:conv}; note that the plateau in the small $m_a$ region corresponds to the parameter space in which this  approximation is applicable.

Finally, we are ready to obtain the number of photons produced in the magnetized region by convoluting \cref{eq:conv} with the flux in \cref{eq:flux2}. This reads
\begin{align}
N_\gamma=T\pi R^2\displaystyle\int\frac{d\Phi_a}{dE_a}(E_a)\mathcal{P}_{a\rightarrow\gamma}\,dE_a\,,
\label{eq:event_number}
\end{align}
where we also included the exposure time, $T$, and the cross sectional area of the magnetized region (cavity), which we assume to be a cylinder of radius $R$ and length $L$. \cref{eq:event_number} is the expression based on which we will discuss sensitivity projection estimates. In passing, let us mention that there is yet another possibility for the detection; ALP can, in principle, decay into $2$ photons within the cylinder; however, since $D\gg L$,  the probability for the decay within the magnetized volume is negligible when compared to the $\mathcal{P}_{a\rightarrow\gamma}$.\\\\

\begin{figure}[t]
    \centering
    \includegraphics[width=0.49 \textwidth, trim= 1 1 1 11, clip]{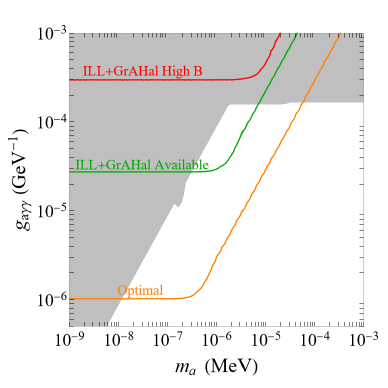}
    \caption{Sensitivity projections in the $m_a$--$g_{a\gamma\gamma}$ parameter space for the considered scenarios where various combinations of reactors and magnets are employed. In gray, we superimpose constraints from existing laboratory experiments, adopted from \cite{AxionLimits}.}
    \label{fig:3}
\end{figure}

\textbf{Experimental setup and sensitivities.}
Before presenting the main results, let us discuss the envisioned experimental setup and background events. For a successful measurement, a photon detection system should be placed right behind the magnetized region. Regarding detectors, one option is to use inorganic scintillators, e.g. NaI[Tl], LaBr3(Ce), with the respective efficiencies in the ballpark of $\mathcal{O}(\%)$ for MeV photon energies \cite{6418383,MOUHTI2018335}. Another option is to use the same material that CAST installed for their solar axion searches -- CdWO4 -- and for this scintillator crystal, the detection efficiencies are in the ballpark of $30-40\%$ \cite{refId0,cryst10060429}. We have included these in our calculations.

We will now make conservative background estimates based on CAST studies for the performance of CdWO4 detector in search for axions with MeV energy \cite{CAST:2009klq}. In \cite{CAST:2009klq}, after applying the cuts, the irreducible background rate is about 1 event per second. Given the period of $T=3$ years that we will typically use in our sensitivity studies, this amounts to $\sim 10^8$ background events. To get the exclusion at 1$\sigma$ level, we would then require the number of signal events to be approximately equal to $\sqrt{10^8}$. Given that the number of ALP-induced photon events scales with $g_{a\gamma\gamma}^4$ (2 powers of $g_{a\gamma\gamma}$ in \cref{eq:cross} and another two in \cref{eq:conv}), this conservative estimate would lead to the reduction of the sensitivity in $g_{a\gamma\gamma}$ by $1$ order of magnitude compared to the ideal case with no backgrounds (where $\mathcal{O}(1)$ signal events suffices for the discovery). We should still stress that \cite{CAST:2009klq} dates back 15 years; to the best of our knowledge, there are no more recent studies by CAST on MeV energy solar axion searches so we are not aware of any more recent improvements with regard to the CdWO4 detectors. However, in this time period, CAST was able to reduce backgrounds in one of their x-ray detectors, Micromegas \cite{Dafni:2008fpa},  (used for lower energy, namely keV axion searches) by two orders of magnitude \cite{GarciaPascual:2015skt}. Assuming such background reduction were achievable for CdWO4 detector as well, the 
sensitivity would be only a factor of $\sim 5$ worse than in the zero background limit. Note also that the cross sectional area of CAST is $14.5 \, \text{cm}^2$
and for one of the proposed GrAHal \cite{Grenet:2021vbb,2022MS&E.1240a2122P} configurations we consider the realization with $R=1.7$ cm, which further improves the situation, given the scaling of the background events with $R^2$.

The main backgrounds considered above arise from cosmic rays and radioactivity of the detector material. In addition to that, the experimental setup we propose in this work may also suffer from background events stemming from the nearby nuclear reactor. Let us, hence, discuss particles that are emitted from the power plants. Photons, that we use to produce ALPs in the reactor core, are effectively absorbed and do not reach the detector. However, neutrons from reactors can travel grater distances and in collision with nuclei (in or near the detector) they could induce the appearance of secondary photons in the detection system. Such background was already under consideration for STEREO experiment \cite{STEREO:2019ztb} near ILL. In Ref.~\cite{kandzia:tel-01796989}, it was shown that neutron-induced background can be effectively suppressed with extra shielding. Reactors are also abundant sources of neutrinos; however, the size of the proposed experimental configuration is such that, given the small neutrino cross sections, neutrino-induced background events are not expected. Given the discussion above, we do not expect reactor-related backgrounds. This can also be verified at particular reactor by comparing reactor-on versus reactor-off photon rates, see Ref.~\cite{PROSPECT:2020sxr}. 


In what follows we will show 95\% CL sensitivity projections ($\chi^2=3.841$) obtained by solving \cref{eq:event_number}, assuming optimistically that all the backgrounds can be removed. Note that, as discussed above, the irreducible backgrounds would in the worst case lead to a factor of few reduction in the sensitivity for $g_{a\gamma\gamma}$. \\

We have considered several different possible scenarios:

\begin{itemize}
\item ALP production at ILL, photon detection at GrAHal, with $B$ and $R$ taken from the \emph{last} line of Table I in \cite{Grenet:2021vbb} ($B=9.5$ T, $R=40$ cm) and  L=$80$ cm. In \cref{fig:3}, this scenario is denoted as ``ILL+GrAHal Available'' as the magnetic field
matches the vanilla setup. The sensitivity is shown by the green line in \cref{fig:3}.

\item ALP production at ILL, photon detection at GrAHal, with $B$ and $R$ taken from the \emph{first} line of Table I in \cite{Grenet:2021vbb} ($B=43$ T, $R=1.7$ cm) and  L=$3.4$ cm~\cite{9714155,2022MS&E.1240a2122P}. In \cref{fig:3}, this scenario is denoted as ``ILL+GrAHal High B''; indeed it was shown that the setup with $B>40$ T is being commissioned at GrAHal. For both this and the previous scenario $D=700$ m and we take $T=3$ years. The sensitivity is shown by the red line in \cref{fig:3}.

\item ALP production at ILL and detection with ILL magnets~\cite{ILLHori,ILLVert} at distance of $D=50$ m. With an ILL magnet of $B=10$ T, $L=318$ mm and $R=19.5$ mm~\cite{ILLVert} we find the reach in $g_{a\gamma\gamma}$ to be reduced by roughly a factor of 2
with respect to ``ILL+GrAHal Available'' projection.

\item ALP production at Bugey and detection with CAST at CERN. The distance between Bugey and CERN is $D\sim 75$ km. Further, $B=9$ T, $L\approx 9$ m and $R=\sqrt{14.5/\pi \,\text{cm}^2}$ \cite{CAST:2017uph}. CAST has been taking data approximately 3 hours per day
(sunrise and sunset) but, if oriented and fixed in the direction of the Bugey reactor, it could be operative throughout the whole day. This kind of realization is imaginable since CAST is approaching its end of data taking and will be superseded by BabyIAXO and ultimately IAXO \cite{IAXO:2020wwp}. The sensitivity that we found for this scenario is $g_{a\gamma\gamma} \gtrsim 10^{-4}$ $\text{GeV}^{-1}$ meaning that a stronger and/or closer magnet would be required to surpass existing laboratory constraints.

\item We also consider the ``Optimal'' configuration where we pair the nuclear power plant with the largest available thermal power, Kashiwazaki-Kariwa power plant near Tokyo ($P\simeq 8.2$ GW), and BabyIAXO magnet at the distance of $D\sim 50$ m. We adopt $R=35$ cm, L=$10$ m, $B=2$ T \cite{IAXO:2020wwp} and assume $T=10$ years of running. The sensitivity is shown by the orange line in \cref{fig:3}.

\end{itemize}

The scenario ``ILL+GrAHal Available'' appears particularly promising; it exceeds laboratory constraints (shown in gray in \cref{fig:3})  for $m_a\gtrsim 10^{-1}$ eV. Despite the smaller $B$ in that case, $R$ is much larger than in ``ILL+GrAHal High B'' and the leading effect hence stems from the angular acceptance.  For ``ILL+GrAHal Available'', the ALP mass for which sensitivity starts ceasing is smaller than for the ``ILL+GrAHal High B'' combination. This can be understood  from ALP conversion formula given in \cref{eq:conv}; sensitivity starts dropping for $m_a^2 L/E_a\gtrsim 1$ and since $L$ is much larger for the available realization, this feature appears at smaller values of $m_a$. 

Finally, the results for the most promising scenario, in which we paired world's strongest nuclear reactor and BabyIAXO magnet, are shown by the ``Optimal'' sensitivity in \cref{fig:3}. Here, both large ALP flux and strong $B$ field are at work making the sensitivity reach $g_{a\gamma\gamma} \approx 10^{-6}\, \text{GeV}^{-1}$. One can infer a large portion of the yet uncovered  parameter space that can be probed. The astrophysical and cosmological constraints are not included in shown constraints. While, admittedly, they would dominate, ALP production in such scenarios can be strongly suppressed, see e.g. \cite{Masso:2006gc,Mohapatra:2006pv,Brax:2007ak,Jaeckel:2006xm}, the studies motivated by the PVLAS anomaly \cite{PVLAS:2005sku}. \\\\
\textbf{Conclusion.}  
In this work we have proposed marrying two very different facilities in order to search for ALPs. Provided that ALPs interact with photons, they can be copiously produced at nuclear reactors via Primakoff scattering. Through the same interaction term in the Lagrangian, ALPs can convert back to photons in the magnetic field. Grenoble is a great location for conducting such an experiment since it has ILL research reactor and, GrAHal, a haloscope platform with strong magnetic field, at only $700$ meters from each other. In addition to that, we also studied several other interesting possibilities, namely combining the commercial power plant Bugey and CAST experiment at CERN as well as the world's strongest nuclear reactor with the next generation axion helioscope. The sensitivity projections found in this work are, for the ALP mass range $m_a \in [10^{-2},10]$ eV, stronger than the existing laboratory constraints. Given this exciting physics potential and the low-cost aspect of the experiment, we envision this work to initiate a brand new program for ALP searches which may potentially lead to new discoveries.\\\\
\textbf{Acknowledgements.}
We would like to thank Rafik Ballou, Juan Collar, Bhupal Dev, Sebastian Ellis, Thierry Grenet, Doojin Kim, Julien Lagoute, David Miller, Pierre Pugnat and Konstantin Zioutas for very helpful discussions. The work of FAA and JQ is supported by the CNRS IN2P3 Master projects A2I, UCMN and by the French National Research Agency (ANR) in the framework of the GrAHal project (ANR-22-CE31-0025). FAA is also supported by the INFN “Iniziativa Specifica” Theoretical Astroparticle Physics (TAsP-LNF) and by the Frascati National Laboratories (LNF) through a Cabibbo Fellowship, call 2022. JQ would also like to thank LAPTh Annecy for hospitality during the final stages of this work. 

\bibliographystyle{JHEP}
\bibliography{bibliography}

\end{document}